# First principles investigations of electronic, magnetic and bonding peculiarities of uranium nitride-fluoride UNF.

Samir F Matar

CNRS, University of Bordeaux, ICMCB. 33600 Pessac. France

Emails: matar@icmcb-bordeaux.cnrs.fr and abouliess@gmail.com

**Abstract:**

*Based on geometry optimization and magnetic structure investigations within density functional theory, unique uranium nitride fluoride UNF, isoelectronic with $UO_2$, is shown to present peculiar differentiated physical properties. Such specificities versus the oxide are related with the mixed anionic sublattices and the layered-like tetragonal structure characterized by covalent like $[U_2N_2]^{2+}$ motifs interlayered by ionic like $[F_2]^{2-}$ ones and illustrated herein with electron localization function graphs. Particularly the ionocovalent chemical picture shows, based on overlap population analyses, stronger U-N bonding versus N-F and d(U-N) < d(U-F) distances. Based on LDA+U calculations the ground state magnetic structure is insulating antiferromagnet with ±2 $\mu_B$ magnetization per magnetic subcell and ~2 eV band gap.*

## 1. Introduction

From the iso-electronic relationship for valence shell states: 2 O ($2s^2$, $2p^4$) ≡ N ($2s^2$, $2p^3$) + F ($2s^2$, $2p^5$), nitride-fluorides of formulation $M^{IV}NF$ type can be considered as pseudo-oxides and isoelectronic with $M^{IV}O_2$ ($M^{IV}$ stands for a generic tetravalent metal). Versus homologous oxides, some relevant physical properties can be expected due to differentiated bonding of M with nitrogen and fluorine qualified as less and more ionic respectively. A small number of tetravalent metal nitride-fluorides exist like transition metal based TiNF [1] and ZrNF [2] on one hand and heavier actinide equiatomic ternaries as ThNF [3] and UNF [4,5] on the other hand. The latter early experimental works in the late 1960's were followed by a review work on nitrides and nitride halides by O'Keefe and Brese in 1992 [6]. Besides the actinide based compounds, only the rare earth ternary CeNCl was evidenced by Ehrlich et al. in 1994 [7]. Recently CeNF [8] was proposed with potential experimental synthesis routes besides full account of physical properties based on extended theoretical works within well established density functional theory DFT [9, 10]. In fact it has been shown in recent decades that this theory with DFT-based methods have allowed not only to explain and interpret experimental results resolved at the atomic chemical constituents scale but also to operate as a predictive tool to propose new stoichiometries with targeted specific properties. As an example of largely investigated systems with BCN pseudo phase diagram, binary carbon nitrides and ternary boron carbon nitrides were identified theoretically with high hardness close to well known ultra-hard diamond and potentially enabled to replace diamond in tooling machinery industry and other abrasive and forage applications [11].

In continuation of our exhaustive investigations of nitride fluorides (cf. [8] and therein cited works) we focus herein on the electronic and magnetic properties of UNF isoelectronic with well known $UO_2$ [12]. UNF is tetragonal and crystallizes in *P*4/*n* space group (SG) which is close to, yet different from, the PbFCl-like one (SG *P*4/*nmm*).The structure is shown in Fig. 1. In spite of its overall three dimensional edifice it can be considered along *c*-tetragonal axis as successions of $U_2N_2$–like motifs separated by $F_2$–like layers, this is supported by the shorter U-N versus U-F distances: 2.29 Å versus 2.61 Å. It will be shown that this structural setup has important influence on the electronic distribution (cf. Fig. 2, *vide infra*).

## 2- Computational methodology

Within DFT we have firstly used the Vienna ab initio simulation package (VASP) code [13,14] for geometry optimization, total energy calculations as well as establishing the energy-volume equations of state. The projector augmented wave (PAW) method [14,15], is used with atomic potentials built within the generalized gradient approximation (GGA) scheme following Perdew, Burke and Ernzerhof (PBE) [16]. This exchange-correlation XC scheme was preferred to the local density approximation LDA [17] one which is known to be underestimating interatomic distances and energy band gaps. The conjugate-gradient algorithm [18] is used in this computational scheme to relax the atoms of the different crystal setups. The tetrahedron method with Blöchl corrections [19] as well as a Methfessel-Paxton [20] scheme was applied for both geometry relaxation and total energy calculations. Brillouin-zone (BZ) integrals were approximated using a special k-point sampling of Monkhorst and Pack [21]. The optimization of the structural parameters was performed until the forces on the atoms were less than 0.02 eV/Å and all stress components less than 0.003 eV/Å$^3$. The calculations were converged at an energy cut-off of 400 eV for the plane-wave basis set with respect to the k-point integration in the Brillouin zone (BZ) with a starting mesh of 6 × 6 × 6 up to 12 × 12 × 12 for best convergence and relaxation to zero strains. The charge density issued from the self consistent calculations can be analyzed using the AIM (atoms in molecules theory) approach [22] developed by Bader. Such an analysis can be useful when trends between similar compounds are examined; it does not constitute a tool for evaluating absolute ionizations. Bader's analysis is done using a fast algorithm operating on a charge density grid arising from high precision VASP calculations and generates the total charge associated with each atom.

From the calculations we also extract information on the electron localization EL at atomic sites thanks to the EL function ELF [23,24] Through normalizing ELF function between 0 (zero localization) and 1 (strong localization) -with the value of ½ corresponding to a free electron gas behavior- enables analyzing the contour plots following a color code: blue zones for zero localization, red zones for full localization and green zone for ELF= ½, corresponding to a free electron gas.

Then for a full account of the electron structure, the site projected density of states (PDOS) and the properties of chemical bonding based on overlap matrix ($S_{ij}$) with the COOP criterion [25] within DFT, we used full relativistic full potential augmented spherical wave (ASW) method [26, 27] with GGA scheme [16].The basis set, limited in ASW method, was chosen as to account the outermost shells to represent the valence states for the band calculations. So that the matrix elements were constructed using partial waves up to $l_{max}$ + 1 = 4 for U and $l_{max}$ + 1 = 2 for N ,O and F. Low energy lying F-2s states (much lower than corresponding O and N 2s states) were considered as core states, i.e. not included in the valence basis set; in the limited ASW basis set they were replaced by 3s states. Self-consistency was achieved when charge transfers and energy changes between two successive cycles were such that $\Delta Q < 10^{-8}$ and $\Delta E < 10^{-6}$ eV, respectively. The BZ integrations were performed using the linear tetrahedron method within the irreducible hexagonal wedge following Blöchl [19].

**3-   Geometry optimization and energy dependent results.**

Table 1 shows the starting experimental and calculated atomic positions and $z_U$ structure parameters for both spin degenerate (NSP: non spin polarized) as well as spin resolved (SP: spin polarized) configurations. SP calculations led expectedly to a magnetization of 4 $\mu_B$ (Bohr magnetons) per unit cell or 2 4 $\mu_B$ per formula unit (FU) which arise from the presence of 2 unpaired electrons in U 5f states of tetravalent uranium. Better agreement with experiment is observed with SP calculations. Note that these calculations merely point out to the trend of developing magnetization from present PAW-GGA calculations, i.e. they do not point to the long range magnetic order or the ground state. In fact $UO_2$ in which uranium is also tetravalent is known to be insulating antiferromagnet in the ground state [28] provided that Hubbard U [29] method is used in further calculations as shown for UNF here below.

Trend to magnetic polarization can be checked as function of volume by establishing the energy-volume equation of state in both NSP and SP configurations. We also verify this for $UO_2$. The NSP and SP E(V) curves are shown in Fig. 3. They all show quadratic behavior with systematic lower SP energy minima. The SP solution is favored for larger volumes but both NSP and SP curves merge together at small volumes or high pressure. The fit of the curves with Birch equation of states (EOS) [29] up to the third order:

$$E(V) = E_0(V_0) + \frac{9}{8}V_0 B_0 [(V_0/V)^{2/3} - 1]^2 + \frac{9}{16} B_0 (B'-4) V_0 [(V_0/V)^{2/3} - 1]^3$$

provides equilibrium parameters: $E_o$, $V_o$, $B_o$ and $B^{'}$ respectively for the energy, the volume, the bulk modulus and its pressure derivative. The obtained values with goodness of fit $\chi^2$ magnitudes are displayed in the inserts of Fig. 3. The equilibrium volumes for both compounds UNF (Table 1) and $UO_2$ come close to experiment: $V(UO_2)$ = 163.73 Å$^3$/cell or 40.93 Å$^3$/FU [30]. The SP volumes are larger than NSP ones and the corresponding energies are lower. Also $\Delta E_{UO2}$(SP-NSP)= -0.23 eV/FU whereas $\Delta E_{UNF}$(SP-NSP)= -0.29 eV/FU meaning that somehow the U-N bond less ionic than U-O prevails. This can be verified from the trend of charge transfer between the two compounds which can be rationalized from the analysis of charge density resulting from the calculations within AIM theory based on Bader's work [31]. Such analysis is particularly relevant when it comes to compare two electronically close compounds as UNF and $UO_2$ here. The results of computed charge changes Q between neutral and ionized elements and resulting overall $\Delta Q$ in the structure are:

UNF:   Q(U) = + 2.29; Q(N) = -1.47; Q(F) = - 0.82.  $\Delta Q$ = ± 2.29.

$UO_2$:   Q(U) = + 2.48; Q(O) = -1.24.  $\Delta Q$ = ± 2.48.

Charge transfer is as expected from U to N O, F with different magnitudes not translating their formal ionizations of "$N^{3-}$", "$O^{2-}$" "$F^{-1}$" as one would expect in the solid state. The overall $\Delta Q$ translating the total ± transfer is smaller for UNF which stresses further the covalent role brought by N through the formation of *U2N2* layers as illustrated below.

Also it is interesting to note the large difference of magnitude of bulk modules $B_0$ pointing to more compressible UNF than UO2. This is partly due to the larger volume of UNF (43.5 Å$^3$/FU) versus $UO_2$ (39.95 Å$^3$/FU) i.e. the larger the volume the more compressible the compound; but this could also be due to the rather layered nature of UNF structure versus tri-dimensional fluorite type $UO_2$

At this point it is interesting to show the 2D-like structure from the point of view of electron localization which is expected to illustrated further the different chemical behaviors of N and F versus U with smaller U-N versus U-F distances observed experimentally. The electron localization EL with the ELF function based on the kinetic energy [32, 33] is used here. In the projections blue, green and red contours represent zero, free electron like and strong localizations respectively. Fig. 3 shows ELF slices along (101) plane with a projection over four adjacent cells. The succession of *U2N2*-like and *F2* like planes along tetragonal *c*-axis. The isolated fluorine is displayed by the blue zones of no localization around it.

## 4- Electronic structure and bonding.

All electrons full potential scalar relativistic ASW calculations were then undertaken for assessing the electronic band structure and qualitative analysis chemical bonding. A comparison between UNF and UO2 was done with spin degenerate (NSP) calculations in order to examine the role of each chemical constituent in the site projected density of states (PDOS) as well as in the chemical bonding. For UNF and $UO_2$ top panels in Fig. 4 show the NSP site projected DOS (PDOS). The zero energy along the *x* axis is with respect to the Fermi level $E_F$ which crosses the lower energy part of the U(5f) states within the valence band (VB). The main part of U(5f) is centered in the empty conduction band (CB) above $E_F$ due to the low filling of 5f states. Nevertheless the crossing occurs at a relatively high PDOS. This is connected with instability of the electronic system in such a spin degenerate configuration of both compounds and with the expected onset of magnetic polarization as shown n next paragraph. Large differences characterize the VB where N(2s) are at -15 eV versus O(2s) at -20 eV; these states show little mixing with uranium itinerant states. Oppositely the hybridization (mixing) between uranium itinerant states and those of N, O p states is identified respectively in the energy windows {-5.5; -3 eV} and {-8; -5.4 eV}; this shift of energies is due to the larger electronegativity of O versus N. In agreement with the observation above on the *U2N2* separated by *F2*–like layers characterizing the structure of UNF (cf. text and Fig. 3), there is little mixing noted between U and F states at -7.5 eV. This aspect should be confirmed from the qualitative analysis of the chemical bonding based on the overlap integral $S_{ij}$ (i and j designate two chemical species) as implemented in the ASW method with the COOP criterion. The relative bonding strengths (U-N versus U-F as well as U-O) is done with the integrated COOP, iCOOP shown in Fig. 4 lower panels. In both panels little bonding can be identified in VB lower part where s-like states a dominant; and significant bonding is found above -10 eV with p-states. Comparing the areas below the iCOOP leads to larger U-N iCOOP versus U-F iCOOP leading to prevailing U-N bonding. The U-O bond in $UO_2$ shows closely similar behavior to U-N albeit with slightly larger iCOOP (note that there are 2 FU in UNF and 1 FU in $UO_2$). Nevertheless U-N iCOOP keeps positive bonding behavior above $E_F$ whereas U-O iCOOP drop rapidly to negative magnitude within CB. This is due to the covalent U-N bond versus rather ionic U-O.

Subsequent spin polarized calculations lead to onset of magnetization in both UNF and $UO_2$ with M=2 $\mu_B$/FU. From Fig. 5 showing the corresponding PDOS along the two spin channels (↑;↓) the integer value is due to the full polarization of electrons in ↑ spin PDOS with a gap appearing in ↓ PDOS with an energy shift between ↑ and ↓ U PDOS signaling the onset of magnetic polarization. The non metal s,p states do not show energy shifts. However the calculation were conducted with plain GGA calculations and it is known that for uranium based compounds as $UO_2$ adding Hubbard U repulsive parameter is needed [33]. With U = 4.1 eV the M=2 $\mu_B$/FU magnetization is reproduced with a small gap opening in ↑ spin PDOS as shown in Fig. 6 top panel. The compound exhibits a magnetic semi-conductor-like behavior. Yet in view of the antiferromagnetic $UO_2$, further calculations assuming two magnetic sub cells, one considered as UP population and the second as DOWN population led ±2 $\mu_B$/subcell and to larger opening of the band gap (~2 eV) with an insulator behavior as shown in bottom panel of Fig. 6. The energy is further lowered by -3.36 eV with respect to plain SP calculations above. Then the ground state of UNF is predicted insulating antiferromagnet.

TABLES

Table 1
Experimental [5] and (calculated) crystal parameters for UNF.
Lattice constants and distances are in Å (1 Å = $10^{-10}$ m).

*P*4/*n*. Origin 1.

N (2*a*) 0 0 0 ; F (2*b*) 0 0 ½; U (2*c*) 0 ½ z.

| UNF | Exp.[8] | NSP | SP |
|---|---|---|---|
| *a* | 3.951 | 3.86 | 3.90 |
| *c* | 5.724 | 5.71 | 5.72 |
| V | 89.35 | 85.1 | 87.0 |
| $z_U$ | 0.2024 | 0.200 | 0.205 |
| d(U-F) | 2.61 | 2.59 | 2.58 |
| d(U-N) | 2.29 | 2.24 | 2.28 |

# Figures

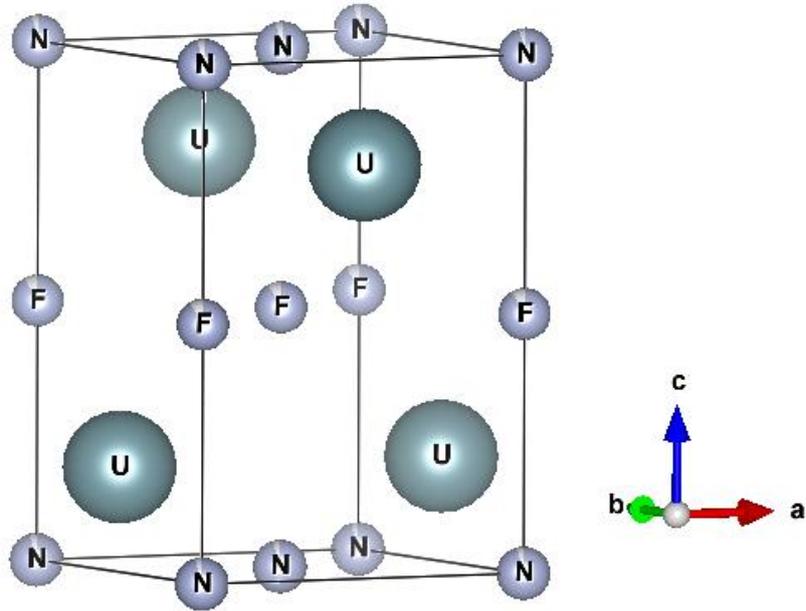

Fig. 1 Tetragonal structure of UNF showing 0 0 ½ planes interlayering {$U_2N_2$}-like blocks.

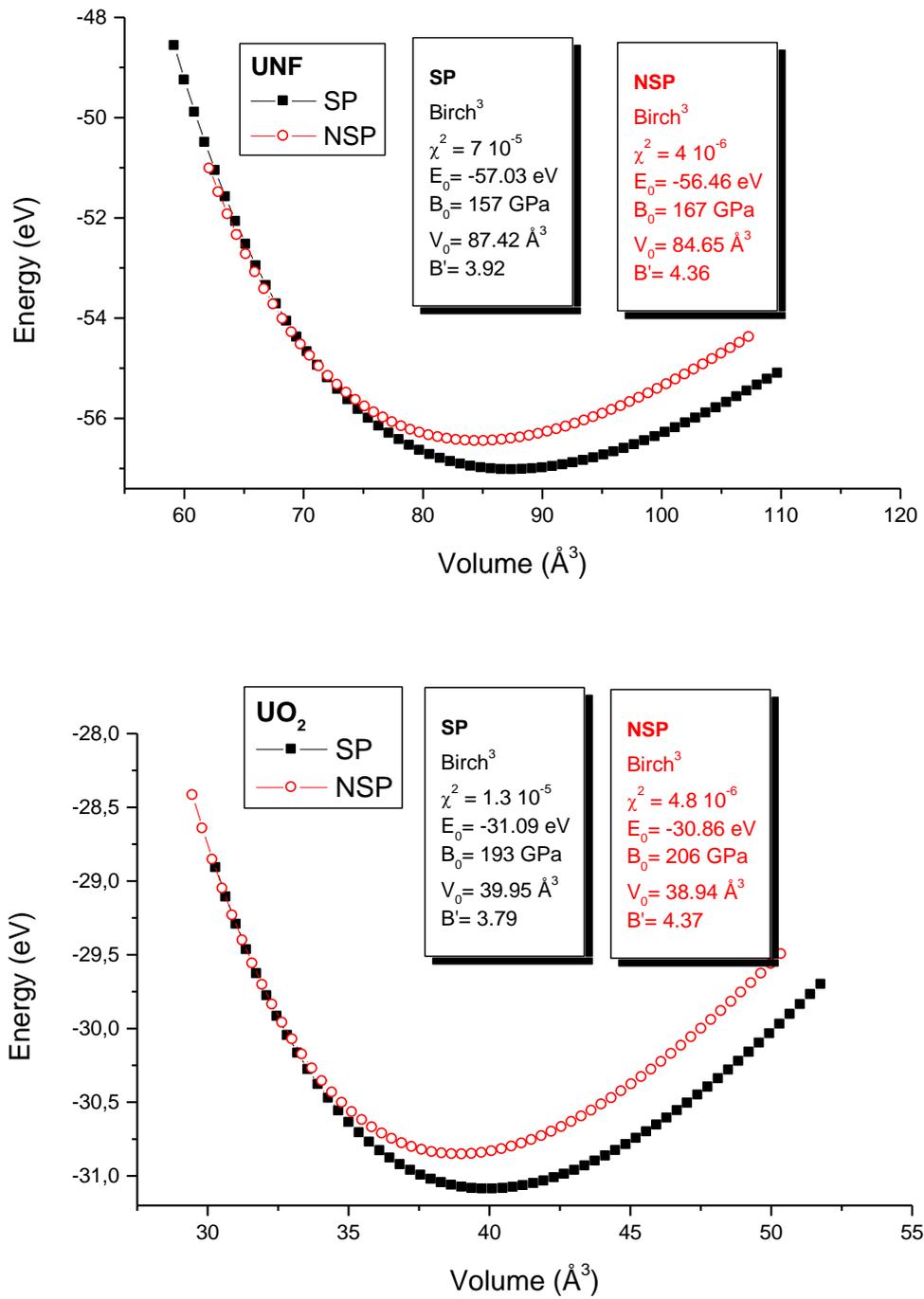

Fig.2 – Energy volume curves and fit values from Birch EOS.

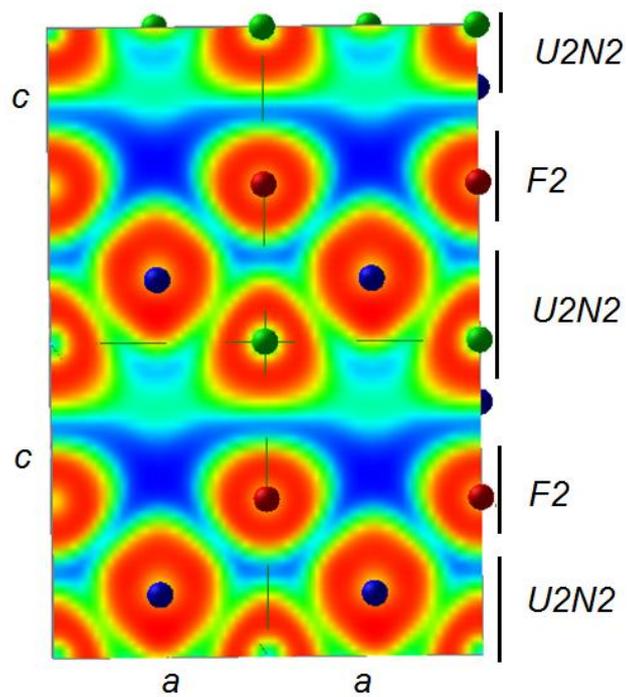

Fig. 3 UNF Electron localization function slice along (101) plane with a projection over four adjacent cells showing the succession of *U2N2*-like and *F2* like planes. Blue, green and red spheres represent U, N and F atoms respectively (see text).

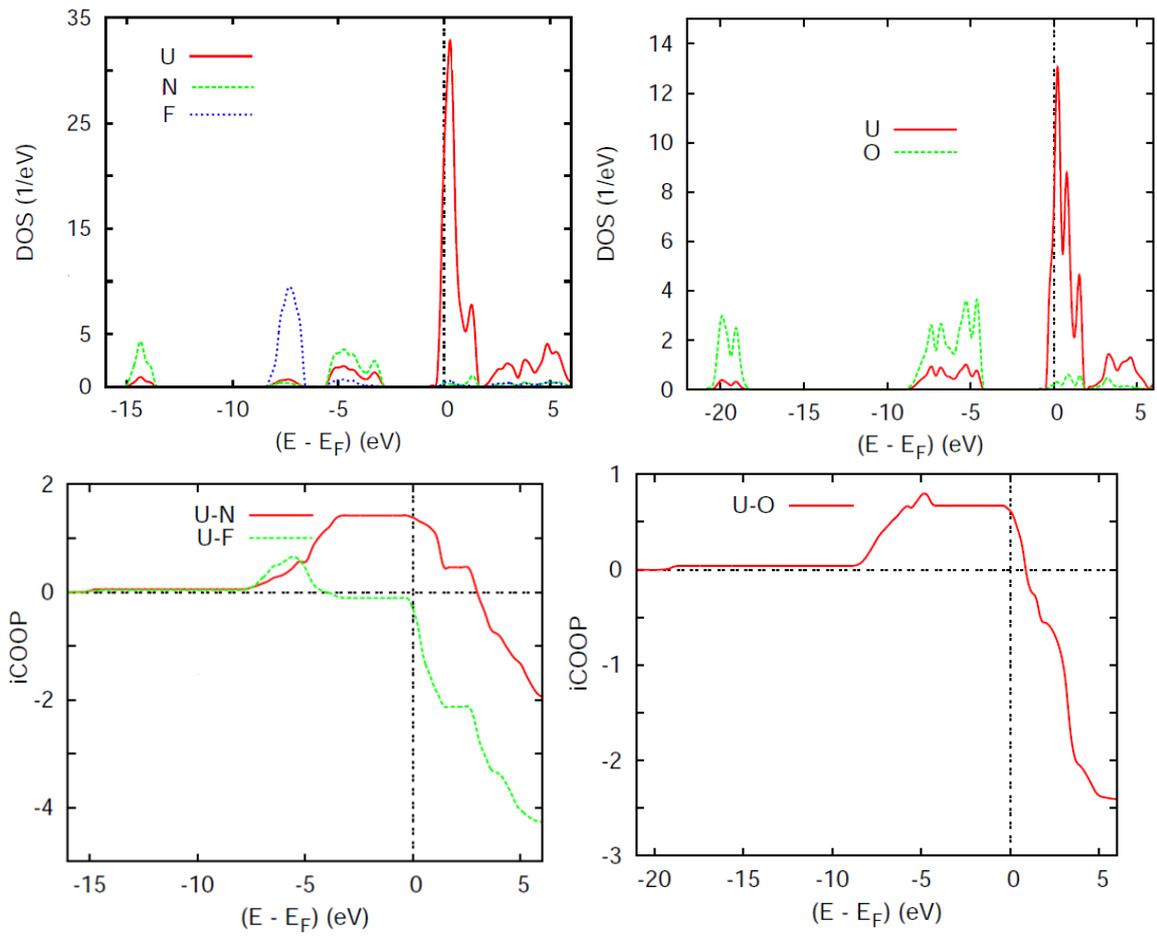

Fig.4 – NSP calculations for UNF (left) and UO$_2$ (right) displaying site projected DOS (up) and chemical bonding from unit less integrated COOP iCOOP (bottom).

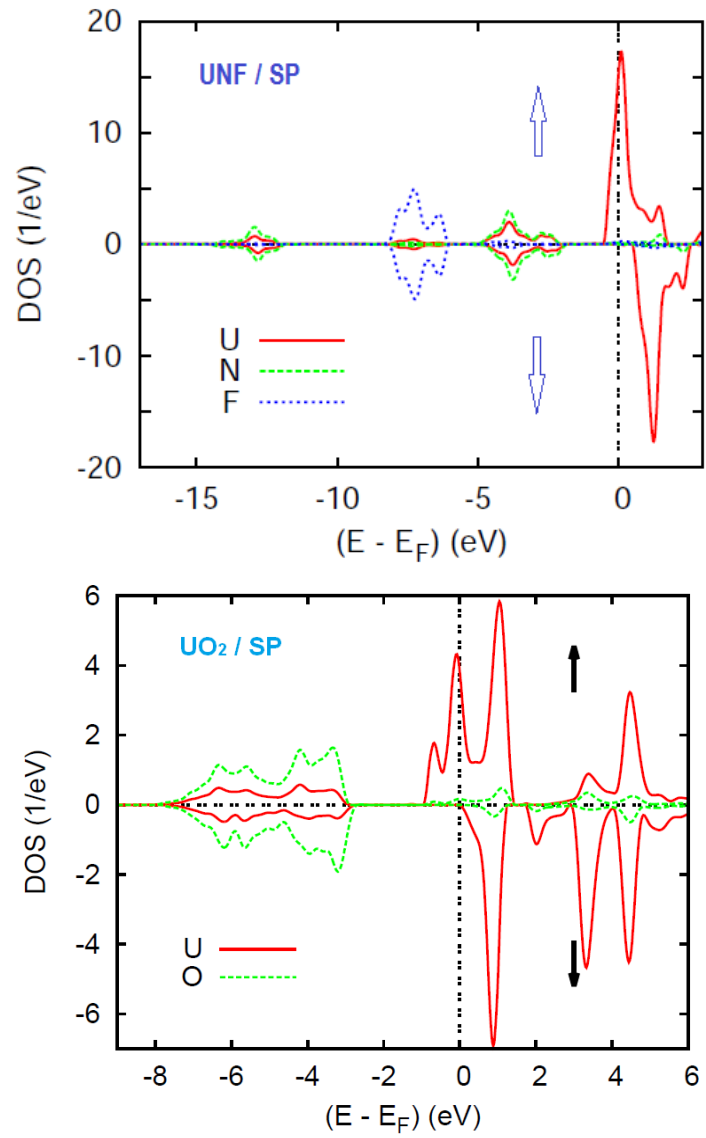

Fig.5 – SP calculations for UNF and $UO_2$ displaying site projected DOS (up) and chemical bonding (bottom).

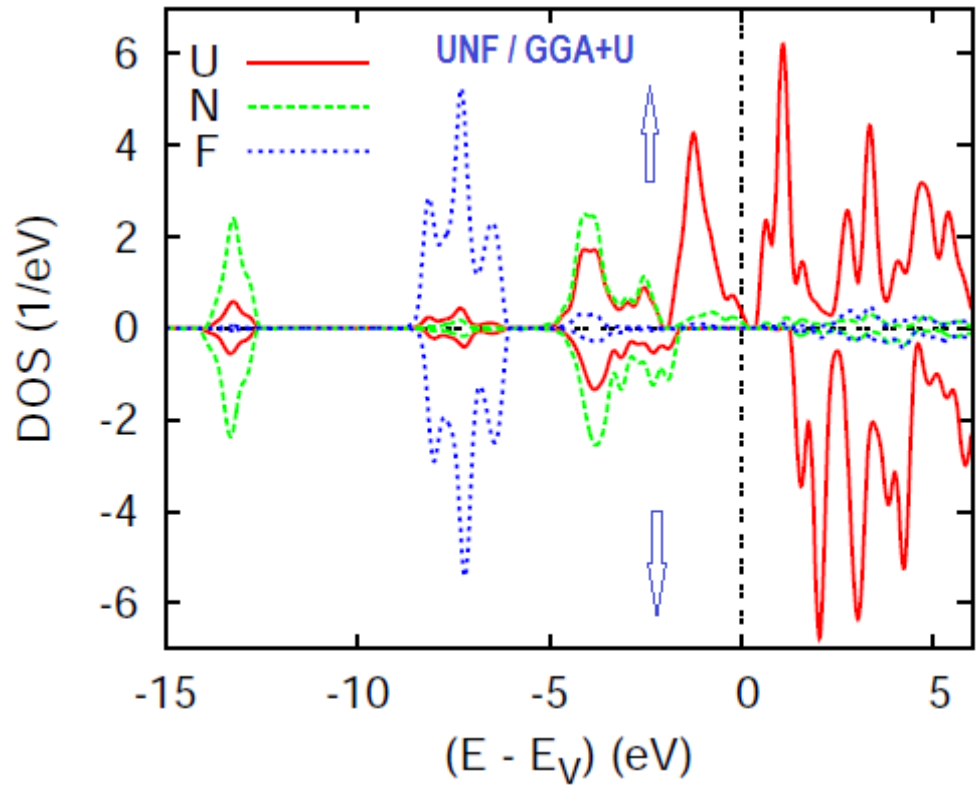

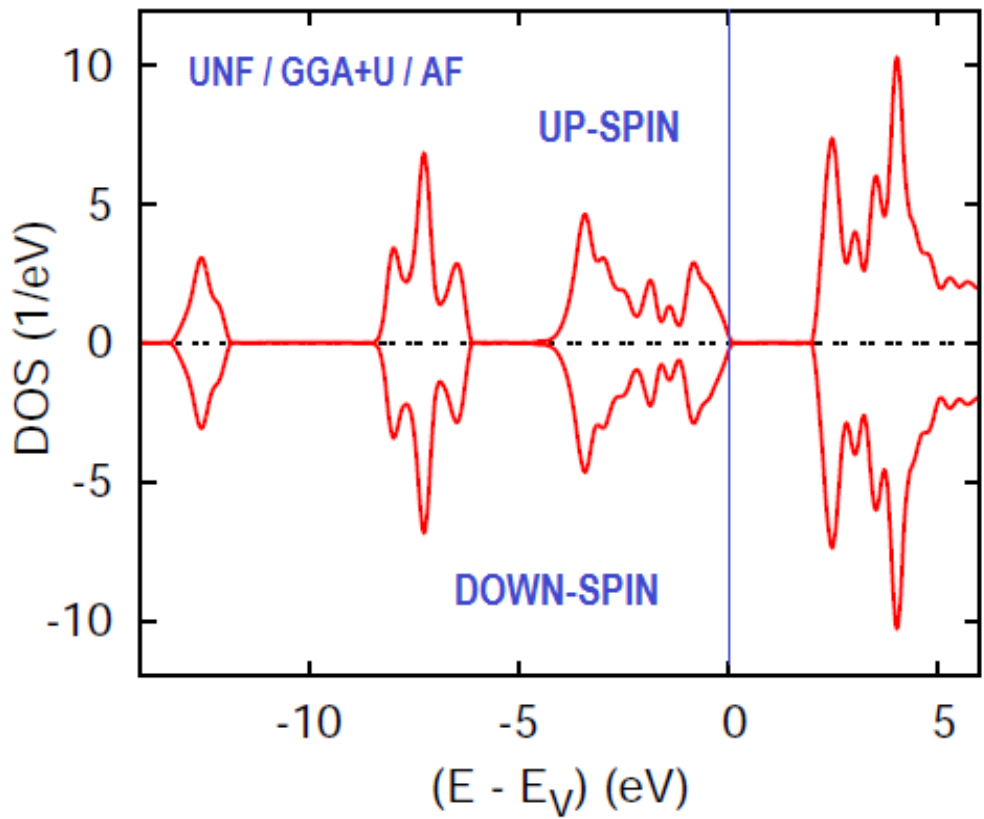

Fig.6 – SP calculations for UNF with GGA+U scalar relativistic calculations.